\documentclass[iop,revtex4]{emulateapj}
\usepackage{natbib}
\usepackage{color}
\usepackage{amsmath}
\usepackage{lscape}
\usepackage{graphicx,setspace,amssymb}

\shorttitle{Fast, Collimated Outflow in the Western Nucleus of Arp\,220}
\shortauthors{Barcos-Mu\~noz et al.}

\begin{document}
\title{Fast, Collimated Outflow in the Western Nucleus of Arp\,220}

\author{Loreto Barcos-Mu\~noz\altaffilmark{1,2}}
\author{Susanne Aalto\altaffilmark{3}}
\author{Todd A. Thompson\altaffilmark{4,5}}
\author{Kazushi Sakamoto\altaffilmark{6}}
\author{Sergio Mart\'in\altaffilmark{1,7}}
\author{Adam K. Leroy\altaffilmark{4}}
\author{George C. Privon\altaffilmark{8,9}}
\author{Aaron S. Evans\altaffilmark{2,10}}
\author{Amanda Kepley\altaffilmark{2}}
\email{loreto.barcos@alma.cl}

\altaffiltext{1}{Joint ALMA Observatory, Alonso de C\'{o}rdova 3107, Vitacura, Santiago, Chile}
\altaffiltext{2}{National Radio Astronomy Observatory, 520 Edgemont Road, Charlottesville, VA 22903, USA}
\altaffiltext{3}{Department of Earth and Space Sciences, Chalmers University of Technology, Onsala Observatory, 439 94 Onsala, Sweden}
\altaffiltext{4}{Astronomy Department, The Ohio State University, 140 W 18th St, Columbus, OH 43210, USA}
\altaffiltext{5}{Department of Astronomy and Center for Cosmology \& Astro-Particle Physics, The Ohio State University, Columbus, OH 43210, USA}
\altaffiltext{6}{Academia Sinica, Institute of Astronomy and Astrophysics, P.O. Box 23-141, Taipei 10617, Taiwan}
\altaffiltext{7}{European Southern Observatory, Alonso de C\'ordova 3107, Vitacura Casilla 763 0355, Santiago, Chile}
\altaffiltext{8}{Instituto de Astrof\'{i}sica, Facultad de F\'{i}sica, Pontificia universidad Cat\'{o}lica de Chile, Casilla 306, Santiago, Chile}
\altaffiltext{9}{Department of Astronomy, University of Florida, 211 Bryant Space Sciences Center, Gainesville, FL 32607, USA}
\altaffiltext{10}{Department of Astronomy, University of Virginia, 530 McCormick Road, Charlottesville, VA 22904, USA}

\begin{abstract}

We present the first spatially and spectrally resolved image of the molecular outflow in the western nucleus of Arp\,220. The outflow, seen in HCN~(1--0) by ALMA, is compact and collimated, with extension $\lesssim$ 120\,pc. Bipolar morphology emerges along the minor axis of the disk, with redshifted and blueshifted components reaching maximum inclination-corrected velocity of $\sim\,\pm$\,840\,km\,s$^{-1}$. The outflow is also seen in CO and continuum emission, the latter implying that it carries significant dust. We estimate a total mass in the outflow of $\geqslant$\,10$^{6}$\,M$_{\odot}$, a dynamical time of $\sim$\,10$^{5}$\,yr, and mass outflow rates of $\geqslant55$\,M$_{\odot}$\,yr$^{-1}$ and $\geqslant\,15$\,M$_{\odot}$\,yr$^{-1}$ for the northern and southern lobes, respectively. Possible driving mechanisms include supernovae energy and momentum transfer, radiation pressure feedback, and a central AGN. The latter could explain the collimated morphology of the HCN outflow, however we need more complex theoretical models, including contribution from supernovae and AGN, to pinpoint the driving mechanism of this outflow.

\end{abstract}


\section{Introduction}

At a luminosity distance of 77\,Mpc (z = 0.018126) and $\mathrm{L_{IR}[8 - 1000 \mu m] \sim 10^{12.2}~L_{\odot}}$ \citep[e.g.,][]{Sanders03}, Arp\,220 is the closest ultraluminous infrared galaxy (ULIRG: $\mathrm{L_{IR} \geq 10^{12}~L_{\odot}}$) and one of the most extreme local star-forming systems. Due to its high infrared luminosity, extreme physical conditions, and proximity, it has been well studied at many wavelengths. It is considered a prototype to understand more distant ULIRGs and is frequently used as a template for starbursts at high redshift.

Arp\,220 is a late-stage merger with two counter-rotating extremely compact nuclei each $<1\arcsec$ in diameter \citep{BM15} and with separation of $\sim1\arcsec$ $\approx369$\,pc \citep[e.g.,][]{Scoville98,Sakamoto99}. Each nucleus harbors gas comparable to the entire content of some galaxies ($\sim 10^9$~M$_\odot$). The large dust column makes Arp\,220 optically thick at mid-IR wavelengths \citep[e.g.,][]{Armus07}, and out to $\sim$100\,GHz \citep{Scoville17,Sakamoto17}. This obscuration has caused a long running debate regarding the nature of the energy source (active galactic nucleus or starburst) in the western nucleus \citep[e.g., see][for an extensive discussion]{Sakamoto17,YH17}.

Molecular outflows have been inferred based on the integrated spectra of both nuclei. Earlier detections include high velocity bluehsifted OH lines \citep{Baan89}, blueshifted absorption in SiO(6--5) \citep{Tunnard15} and HCO$^{+}$\,(3--2) \citep{Martin16}, and P-Cygni profiles of HCO$^{+}$ \citep{Sakamoto09}, H$_{2}$O, H$_{2}$O$^{+}$, and OH$^{+}$ \citep{Rangwala11,GA12}. Recent high resolution ($\sim$0$\farcs$1) observations of HCN, HCO+, and SiO~(2--1) \citep{LBM_phd} also exhibit P-Cygni profiles from the central beam of each nucleus.

This spatially unresolved line emission/absorption provides kinematic information on the outflow, but little morphological information. Continuum emission at 150\,MHz and 104\,GHz along the minor axis of the western nucleus has also been linked to an outflow \citep{Varenius16,Sakamoto17}. This elongated continuum emission suggests the outflow's spatial structure, but does not offer kinematic information. 

Here, we present the first spatially and spectrally resolved image of the fast collimated molecular outflow in the western nucleus of Arp\,220. We use ALMA to obtain resolved spectroscopy of the outflow at 0$\farcs$1 resolution ($\sim$40\,pc) and detect it both HCN~(1--0) and CO~(1--0). Throughout this paper, we adopt H$_{0}$=73\,km\,s$^{-1}$\,Mpc$^{-1}$, $\mathrm{\Omega_{vacuum}=0.73}$ and $\mathrm{\Omega_{matter}=0.27}$, with velocities corrected to the cosmic microwave background frame.

\section{ALMA Observations of Arp\,220} 
\label{sec:obs}

We observed Arp\,220 in three sessions between 2015 October 15 and 2015 October 29 using the Atacama Large millimeter/sub-millimeter Array (ALMA) (PI: L. Barcos-Mu\~noz). The observations targeted the high critical density transitions HCN~(1--0) and HCO$^{+}$~(1--0), their optically thin isotopologues H$^{13}$CN~(1--0) and H$^{13}$CO$^{+}$~(1--0), and the shock tracer SiO~(2--1). In this letter, we focus on the HCN emission detected at 0$\farcs$1 resolution. The full data set will be presented in a future paper.

The total integration time on source was 2 hours. The maximum baseline used was $\approx 11$~km, with the largest recoverable angular scale of 1.1$\arcsec$. We configured the correlator to process four spectral windows, each 1.875\,GHz wide. After on-line smoothing, the spectral resolution near our line of interest was 13.6\,km\,s$^{-1}$.

\begin{figure}[tbh]
\centering
\includegraphics[scale=0.24]{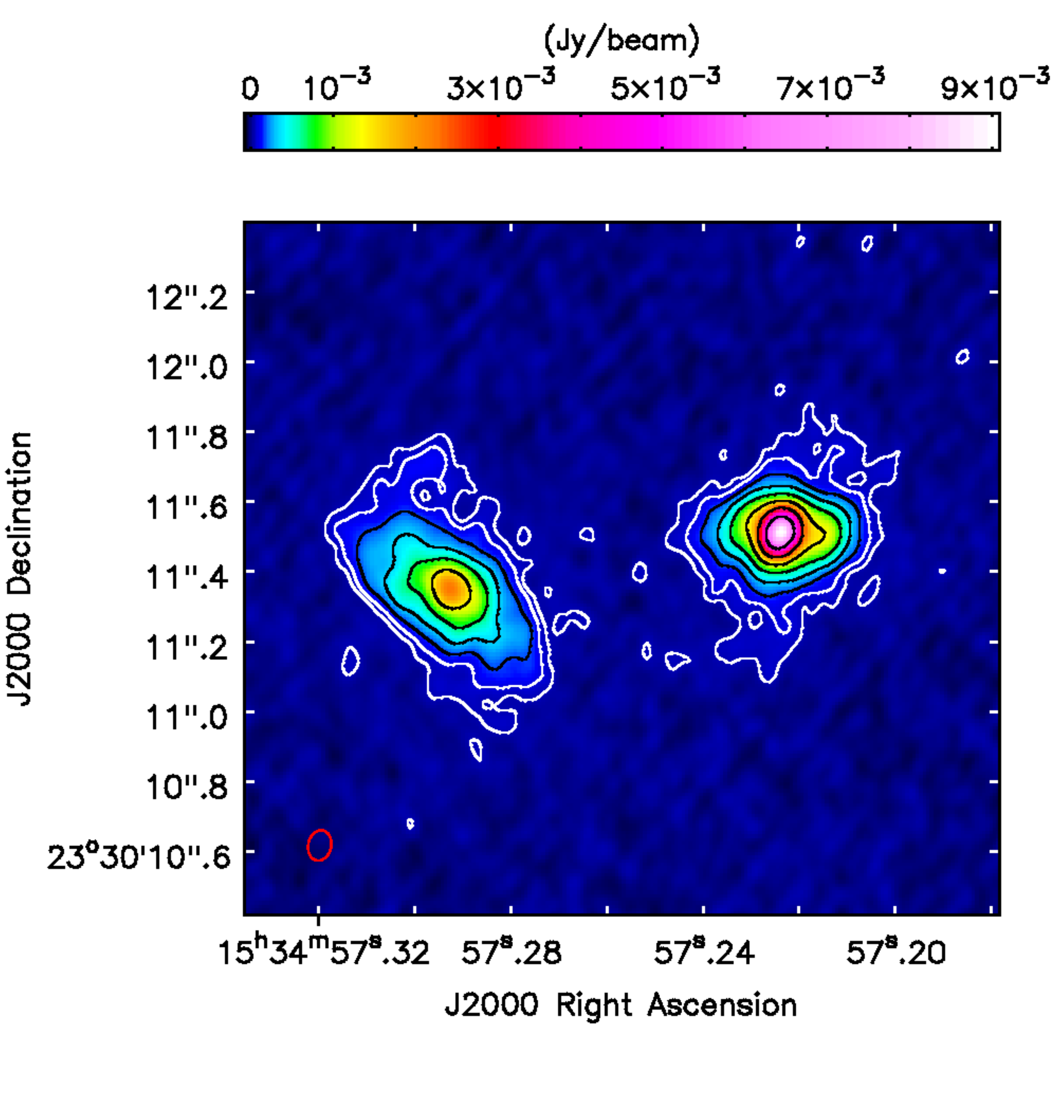}
\caption{ALMA 92.2\,GHz (3.3\,mm) continuum image of Arp\,220 with contours stepped by factors of $2$ beginning at $3\sigma_{\rm 92GHz}$. The FWHM of the restoring beam is 0$\farcs09\times0\farcs07$, P.A.$\approx-13^{\circ}$ (lower left, red). \label{fig:fig1}}
\end{figure}

The data reduction, imaging and analysis were carried out using the Common Astronomy Software Application package \citep[CASA;][]{McM07}. We first ran the observatory-provided reduction script. Then, we identified channels with negligible contamination by spectral lines and produced an image of the 92.2\,GHz (3.3\,mm) continuum emission, which we iteratively self-calibrated. Figure \ref{fig:fig1} shows the resulting 3.3\,mm continuum image. The rms and peak intensity of the final image are $\mathrm{\sigma_{92GHz}=14.4\mu Jy~beam^{-1}}$ (0.36\,K) and 9.1\,mJy\,beam$^{-1}$ (228\,K). The size of the synthesized beam is 0$\farcs09 \times 0\farcs07$, position angle (P.A.) $\mathrm{\approx-13^{\circ}}$.

We applied the self-calibration derived from the continuum to the line data. Then, we subtracted the continuum and imaged the full spectral window containing the HCN~(1--0) line (rest frequency 88.6316\,GHz). Finally, we smoothed the cube to a common angular resolution for HCN~(1--0) of 0$\farcs12\times0\farcs09$, P.A.$\approx-6.7^{\circ}$. The median rms noise per channel is $\sigma_{\rm HCN}$=0.35\,mJy\,beam$^{-1}$ (5\,K).

\section{Results}
\subsection{Fast collimated HCN outflow in the Western Nucleus of Arp\,220}
\label{sec:hcn_outflow}

\begin{figure*}[tbh]
\centering
\includegraphics[scale=0.7,angle=-90]{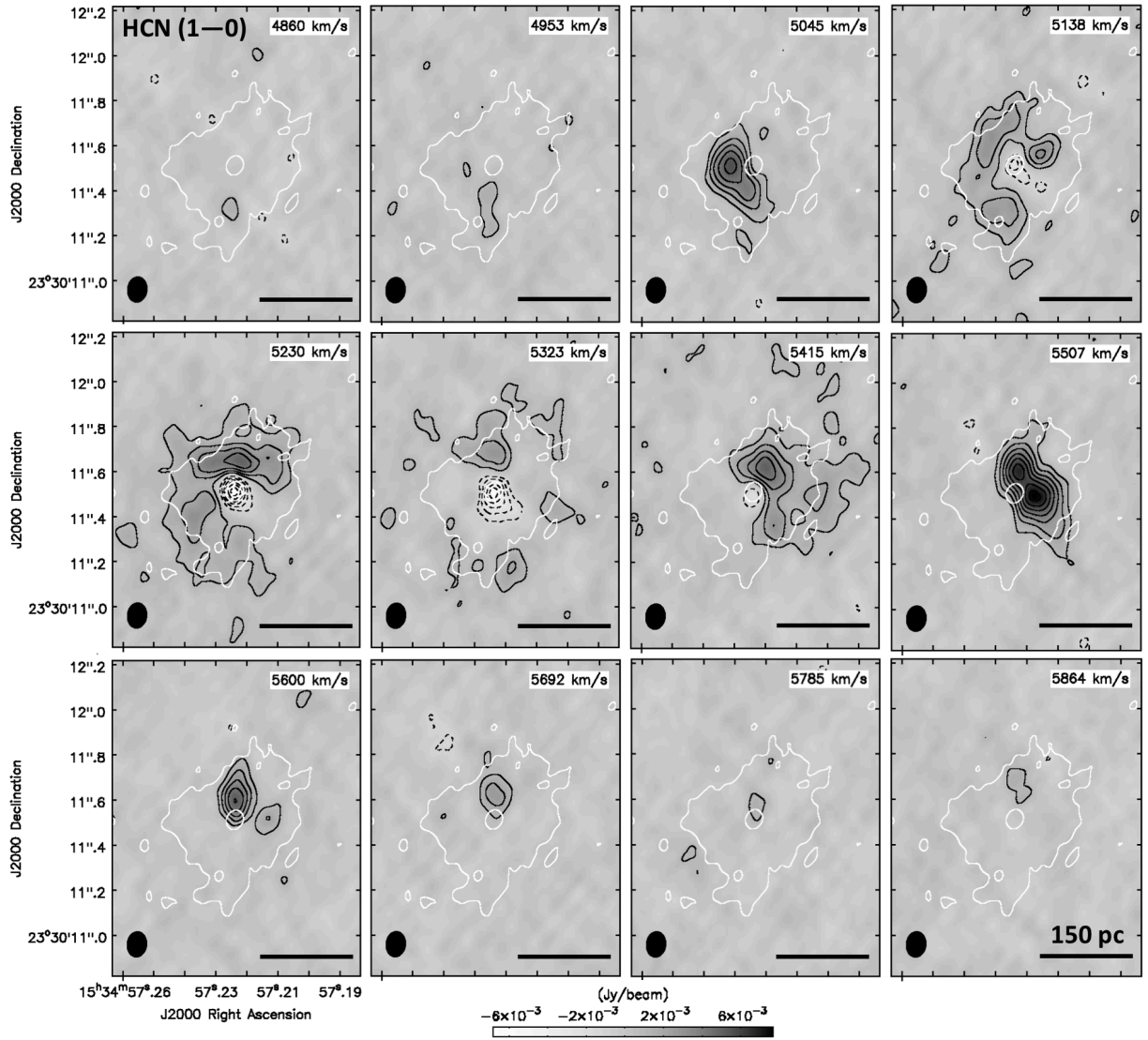}
\caption{Channel maps of HCN~(1--0) emission (grayscale map, see color bar at the bottom) from the western nucleus of Arp\,220. Channels are separated by $\mathrm{7\times13.6~km~s^{-1}}$, except for the last channel which is separated by $\mathrm{6\times13.6~km~s^{-1}}$. The first and last channels are the most blueshifted and redshifted channels showing emission with $S/N>=3$ from the outflow. Black contours begin at $S/N=3$ and are stepped by factors of $3$, with negative contours drawn as dashed lines.  White contours show the $S/N=3$ and $S/N=400$ contours for the 92\,GHz continuum emission (see Fig.~\ref{fig:fig1}). \label{fig:fig2}}
\end{figure*}

Figure \ref{fig:fig2} shows channel maps of HCN~(1--0) emission. In addition to emission from the rotating disk, we observe higher velocity emission perpendicular to the major axis of the disk which is both blueshifted and redshifted relative to the disk. We observe emission having $S/N>2.5$, in consecutive channels, up to blueshifted velocities of $\mathrm{-510~km~s^{-1}}$ and redshifted velocities of $\mathrm{540~km~s^{-1}}$, towards the south and north, respectively. We identify this emission as tracing a bipolar outflow from the disk. We construct a moment zero map of the outflow over the velocity ranges of $\mathrm{270~km~s^{-1} < v-v_{sys} < 540~km~s^{-1}}$ and  $\mathrm{-510~km~s^{-1} < v-v_{sys} < -370~km~s^{-1}}$ where the lower velocity bounds are selected to avoid contamination from the rotating disk. We assume $\mathrm{v_{sys} = 5355 \pm 15~km~s^{-1}}$ for the western nucleus \citep{Sakamoto09}. Note that the low velocity component of the outflow somewhat overlaps the disk, and separating the two remains a source of uncertainty.

\begin{figure*}[tbh]
\centering
\includegraphics[scale=0.6]{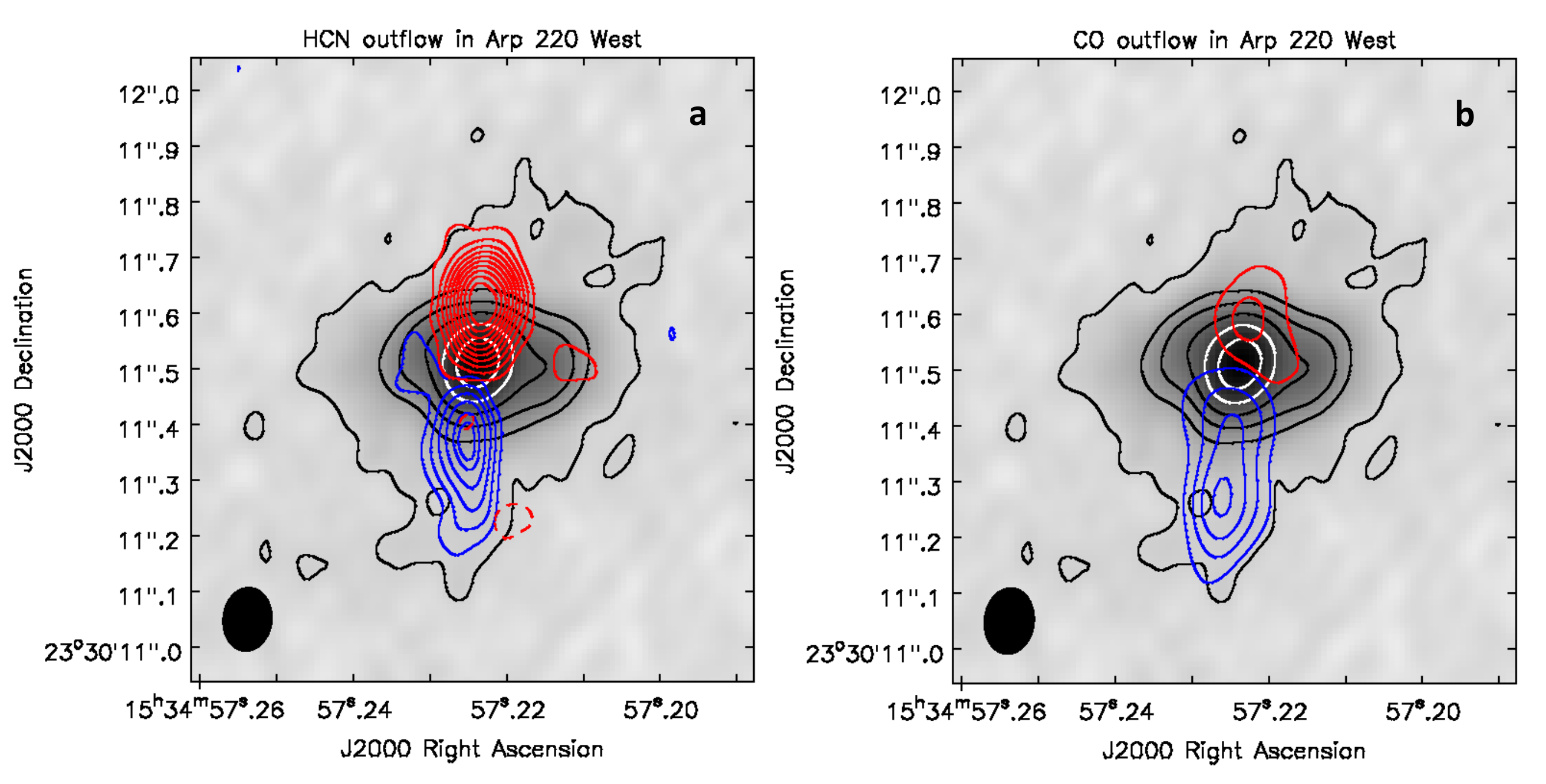}
\includegraphics[scale=0.65]{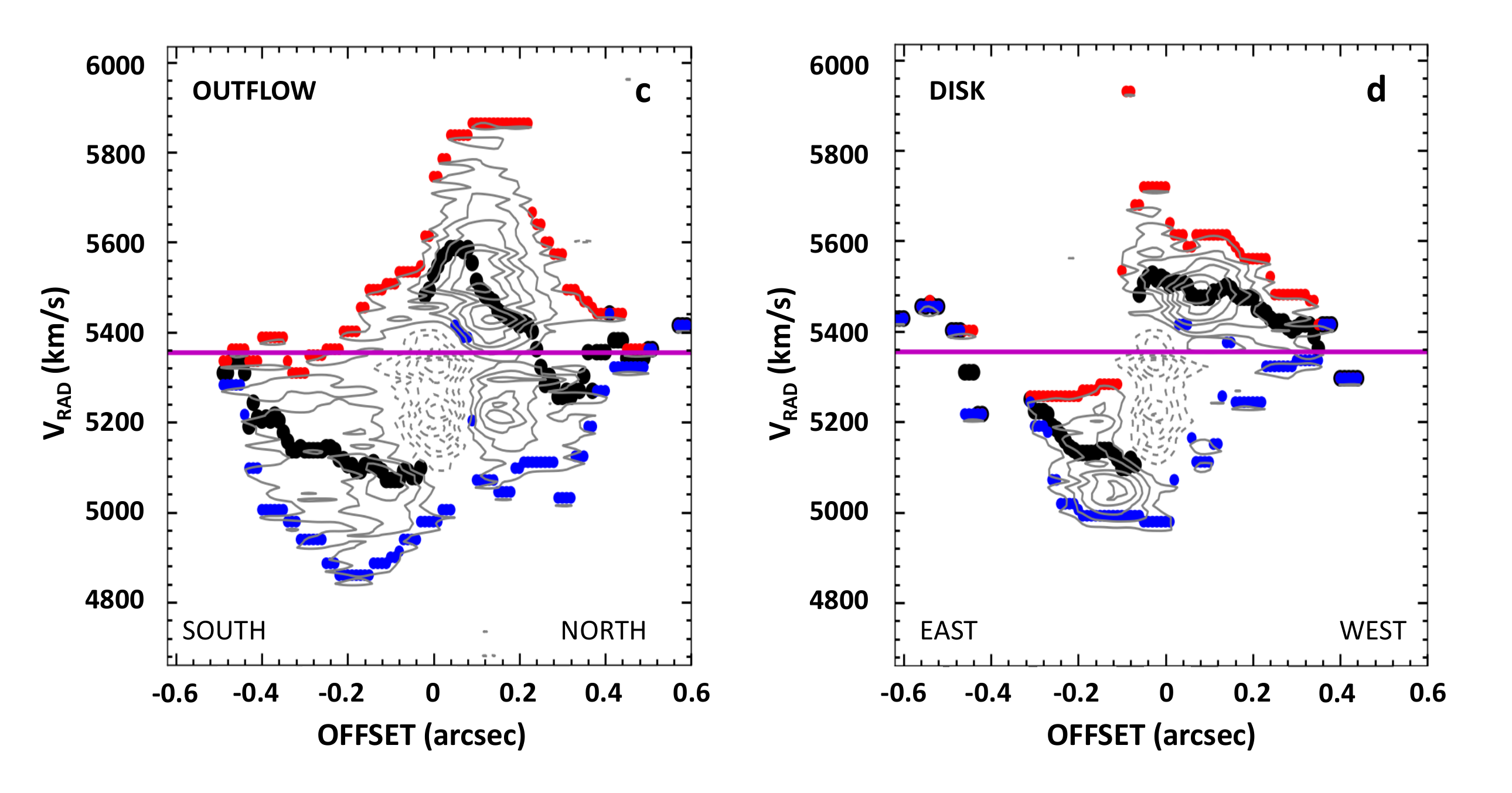}
\caption{{\it (Top Row)} Greyscale: 92\,GHz continuum image of the western nucleus of Arp\,220 with its black (3,24,48 and 96$\times\sigma_{\rm 92GHz}$) and white (192, 384$\times\sigma_{\rm 92GHz}$) contours. Beam size, 0$\farcs$12$\times$0$\farcs$09, of line emission shown in the bottom left corner of each panel in black. {\it (a)} The red contours (-3,3,5,7,9,11,13,15,17,19,21$\times\sigma_{\rm rms}$, where $\sigma_{\rm rms}$ = 23 mJy beam$^{-1}$ km s$^{-1}$) show the HCN(1--0) emission of the integrated intensity map of redshifted channels, while blue contours (3,5,7,9,11,12$\times\sigma_{\rm rms}$, where $\sigma_{\rm rms}$=17\,mJy\,beam$^{-1}$\,km\,s$^{-1}$) show the integrated intensity emission of blueshifted channels. {\it (b)} Moment zero map contours of high velocity channels of CO emission. The red (3 and 5$\times\sigma_{\rm rms}$) and blue contours (3,5,7,9$\times\sigma_{\rm rms}$), where $\sigma_{\rm rms}$=30\,mJy\,beam$^{-1}$\,km\,s$^{-1}$, show the CO~(1--0) emission of the integrated intensity map of the redshifted and blueshifted emission, respectively. {\it (Bottom Row)} HCN position-velocity diagrams of the outflow and the rotating disk of the western nucleus of Arp\,220. {\it (c)} PV diagram from the outflow integrated over a slit 11\,pc wide along a P.A.\ of 173$^{\circ}$. Grey contours are in steps of $\pm$3$\times$2$^{n}\sigma_{\rm HCN}$, where n = 0,1,2,3... . We highlight the minimum (blue), median (black), and maximum (red) velocity across the extent of the slit. The magenta line indicates $\mathrm{v_{sys} = 5355 \pm 15~km~s^{-1}}$. {\it (d)} Same as in the {\it Left} panel but for the disk along a slit 40\,pc wide and a P.A.\ of 80$^{\circ}$.\label{fig:fig3}}
\end{figure*}

The top panels of Figure \ref{fig:fig3} present the first high angular and spectral resolution image of the molecular outflow in the western nucleus of Arp\,220. In the top-left panel, blue and red contours show integrated intensity of the two high velocity HCN lobes. Note that some contamination from the rotating disk along the east-west direction persists. Based on the positions of the integrated emission peak of the two lobes, we calculate a P.A. of about 173$^{\circ}$ for the outflow. This is almost perpendicular to the P.A.$\sim80^{\circ}$ determined for the disk based on the 33\,GHz and 104\,GHz continuum \citep{BM15,Sakamoto17}.

The outflow accelerates away from the disk. In panel c of Figure \ref{fig:fig3}, we show the position velocity diagram (PVD) of the outflow along a 11\,pc (0$\farcs$03) wide slit with a P.A. of 173$^{\circ}$. We highlight the minimum (blue), median (black), and maximum (red) emission velocities at each step along the outflow. From these minimum and maximum velocity profiles, we see that the outflow velocity increases as one moves away from the nucleus, reaching a maximum at about 0$\farcs$2. 

For comparison, panel d of Figure \ref{fig:fig3} shows the PVD of the disk along a slit 40\,pc (0$\farcs$11) wide at the major axis P.A. of 80$^{\circ}$. The shape of the PVD matches a rotating disk whose velocity decreases outwards.

Both panels show strong absorption towards the nucleus itself. For further discussion of HCN and HCO$^{+}$ absorption in a higher J transition, see \citet{Martin16}.

\subsection{Further Evidence for the Outflow}
\label{sec:evidence}

To confirm the features seen in HCN, we produced a similar map of CO emission using ALMA observations presented by \citet{Scoville17} and \citet{Sakamoto17}. We beam-matched the CO to the HCN and selected high velocity channels in the same way that we did for HCN. For CO, the redshifted emission spans $\mathrm{260~km~s^{-1} < v-v_{sys} < 370~km~s^{-1}}$ and the blueshifted emission covers $\mathrm{-520~km~s^{-1} < v-v_{sys} < -370~km~s^{-1}}$. This range of velocities differs from that for HCN mainly due to the absence of CO emission at $\mathrm{v-v_{sys} > 370~km~s^{-1}}$.

Panel b of Figure \ref{fig:fig3} shows the integrated blueshifted and redshifted CO emission. The outflow is clearly present in CO emission, but it is less pronounced than in HCN. The southern lobe is somewhat more spatially extended in CO than in HCN. It is also brighter than the northern lobe in CO, opposite to what we observe in HCN. 

With our current data set, the physical driver of the different CO and HCN structure, especially in the north lobe, remains unclear. Resolved observations of high $J$ transitions, isotopologues, and other species will help resolve whether this is primarily, a chemical, excitation, or radiative transfer effect. Inspection of the HCO$^{+}$(1--0) and SiO(2--1) emission from our ALMA observations suggests the presence of the outflow, but at lower significance than the HCN and CO data presented here. There is also evidence for formaldehyde and OH megamasers along the outflow direction \citep{Rovilos03,Baan17}, suggesting that the wind and masers are physically related.

The outflow can also be seen in dust continuum. In Figure \ref{fig:fig1}, we observe extended emission along the direction of the outflow, which is even clearer at 104\,GHz (Fig.~4 of \citealt{Sakamoto17}). Combining our image with their data, we created a spectral index map between 92 and 104\,GHz. We match the resolution and astrometry of the two images and then masked them at $S/N>5$ for the 104\,GHz image. 

\begin{figure*}[tbh]
\centering
\includegraphics[scale=0.65]{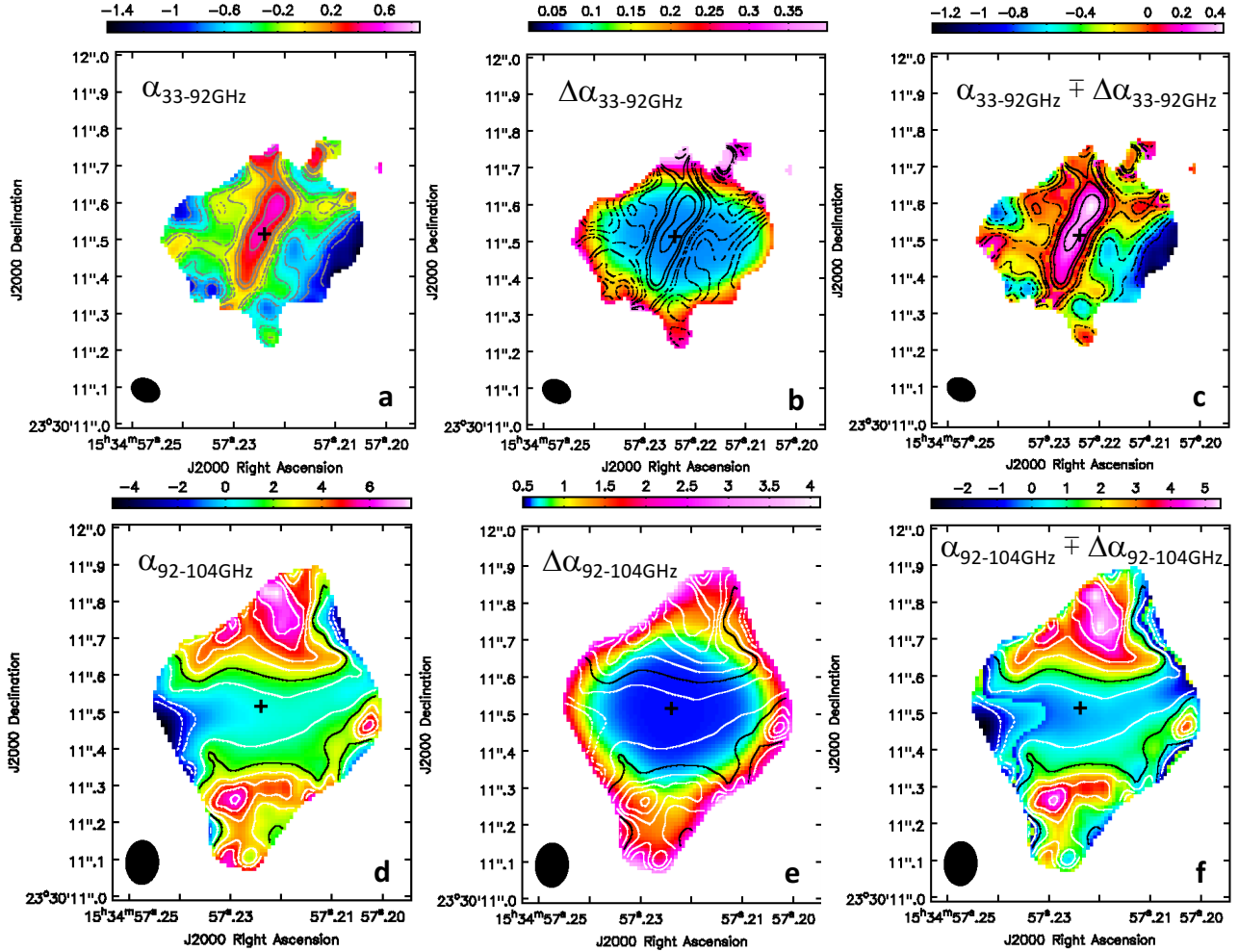}
\caption{Spectral index maps of the western nucleus of Arp\,220. {\it (Top Row)} Spectral index map between 32.5\,GHz and 92.2\,GHz (a), its uncertainty map (b), and a spectral index map indicating lower (upper) limits for positive (negative) spectral indices (c). The resolution in these maps is 0$\farcs$09$\times$0$\farcs$07, (filled black beam in bottom left corner). The contours are $-$1, $\pm$0.8, $\pm$0.6, $\pm$0.4, $\pm$0.2 and $\pm$0.1, with negative contours in dashed lines. {\it (Bottom Row)} Same as top row, but for 92.2 GHz and 104.1 GHz. The white solid contours indicate spectral indices of 1,3,4,5 and 6, and white dashed contours indicate spectral indices of $-2$ and $-1$. The solid black contour corresponds to a spectral index of 2. The resolution in these maps is 0$\farcs$12$\times$0$\farcs$09. In all panels, the black cross indicates the location of the brightest pixel at 92\,GHz of the western nucleus of Arp\,220.\label{fig:fig4}}
\end{figure*}

Panel d of Figure \ref{fig:fig4} shows the resulting map. The extended emission to the north and south has $\alpha \geq 2$ indicating that dust emission likely represents the dominant continuum component. The regions closer to the center show a mixture of synchrotron, free-free, and dust emission. For comparison, panel a of Figure \ref{fig:fig4} shows the spectral index map between 33 and 92\,GHz. Between these two frequencies we expect mostly negative or flat spectral indices coming from synchrotron and free-free emission. Instead, we observe a collimated positive spectral index along the north-south direction. The P.A. of this feature, $\sim$ 153$^{\circ}$, differs only slightly from the P.A. of the HCN outflow, and we attribute this feature to dust associated with the outflow\footnote{Alternatively, a positive spectral index between 33 and 92 GHz may indicate optically thick free-free emission, however the brightness temperatures needed for this, $\sim10^{4}$K, are not observed.}. For both cases, we also provide the spectral index error map (middle column), and a spectral index map with the lower (upper) limit for positive (negative) spectral indices (right). From the latter we observe that the positive spectral indices features along the direction of the outflow, remain positive when including the uncertainties.

\subsection{Properties of the Outflow}
\label{sec:properties}

\underline{\it Luminosity:} Integrating over the $S/N>3$ contours in the integrated intensity map (Figure \ref{fig:fig3}), we find an HCN luminosity of $L_{\rm HCN}\simeq2.1\times10^{7}$\,$\mathrm{K~km~s^{-1} pc^{2}}$ associated with the northern lobe and $8.9\times10^{6}$\,$\mathrm{K~km~s^{-1} pc^{2}}$ for the southern lobe. For CO, we find $L_{\rm CO}\simeq3.1\times10^{6}$\,$\mathrm{K~km~s^{-1} pc^{2}}$ (north) and $9.9\times10^{6}$\,$\mathrm{K~km~s^{-1} pc^{2}}$ (south). These luminosities are lower limits because the low velocity components of the outflow remain confused with the disk.

\underline{\it Velocity:} The maximum velocities detected in HCN are $+509$ (north) and $-495$\,km s$^{-1}$ (south). Assuming that the outflow is perpendicular to the disk (inclination 53.5$^{\circ}$; \citealt{BM15}), these reflect true outflow velocities of $+856$ and $-832$\,km\,s$^{-1}$ (see Table \ref{tbl-1}).

\underline{\it Mass:} The amount of mass implied by this HCN luminosity is highly uncertain. \citet{Gao_HCN04} proposed a conversion factor $\mathrm{\alpha_{HCN}= 10 M_{\odot} (K~km~s^{-1} pc^{2})^{-1}}$ to translate from luminosity to the mass of dense ($\gtrsim 3 \times 10^4$~cm$^{-3}$) gas \citep[but this is quite uncertain, see][]{Leroy17}. The appropriate $\alpha_{\rm HCN}$ should be lower in massive star forming environments because the gas should be hotter and the HCN abundance should be elevated \citep{Gao_HCN04,GB12,Aalto15}. Both appear to be the case in the western nucleus of Arp 220. \citet{Tunnard15} found an excitation temperature of $T_{\rm ex}\sim$150 K and \citet{GA12} found HCN abundance of 10$^{-6}$ compared to the $2 \times 10^{-8}$ adopted by \citet{Gao_HCN04}.

We calculate a lower limit for $\mathrm{\alpha_{HCN}}$ in dense gas assuming local thermodynamic equilibrium (LTE) and low optical depth. Assuming a background temperature of T$_{\rm bg}=$ 2.73 K, we obtain $\mathrm{\alpha_{HCN}\geqslant 0.24 M_{\odot} (K~km~s^{-1} pc^{2})^{-1}}$, following $\mathrm{\alpha_{HCN} \approx 0.0015T_{ex} + 0.015}$, for $\mathrm{10 K \leq T_{ex}\leq 200K}$. If we use instead T$_{\rm bg}=$ 50 K \citep[closer to the value found by][]{Tunnard15}, we obtain $\mathrm{\alpha_{HCN} \approx 0.001T_{ex} + 0.21 \approx 0.36}$, for $\mathrm{100 K \leq T_{ex}\leq 200K}$. Using $\mathrm{\alpha_{HCN}\geqslant 0.24 M_{\odot} (K~km~s^{-1} pc^{2})^{-1}}$, we obtain $M_{\rm H_2}\geqslant 4.9\times10^{6}$\,M$_{\odot}$ and $2.1\times10^{6}$\,M$_{\odot}$ for the north and south lobes, respectively. These lower limits agree with what we obtained trying to match the observed brightness temperature by varying the column density, assuming LTE, and using the size of the lobes (see below) and an H$_{2}$ volume density of $2.1\times10^{6}$\,cm$^{-3}$ \citep{Tunnard15}.

Note that if the outflow has a reasonably high ionization fraction, excitation by collisions from electrons may become important. Because electrons are more effective at collisional excitation than H$_2$ \citep{Faure07}, this would have the effect of lowering the critical density for HCN emission, in which case the optically thin $\alpha_{\rm HCN}$ calculated above might apply to all of the gas, not only dense gas in LTE. A similar case has been considered in NGC 253 by \citet{Walter17} and could also be the explanation of why we see HCN emission from lower density gas, $\sim$500 cm$^{-3}$, in galactic regions \citep{Kauffmann17,Pety17}.

If we instead use the CO luminosity and a CO ULIRG-like conversion factor of $\mathrm{\alpha_{CO}= 0.8 M_{\odot} (K~km~s^{-1} pc^{2})^{-1}}$, we derive $M_{\rm H_2}\geqslant2.5\times10^{6}$\,M$_{\odot}$ and 8.0$\times$10$^{6}$\,M$_{\odot}$ for the north and south lobes, respectively. Note that the CO estimates of the outflow lobe masses are also lower limits since so is L$_{\rm CO}$.

We also tried using dust emission at 104\,GHz from the outflow to estimate the gas mass, however we obtained values close to the dynamical mass of the entire nucleus. Uncertainties in this calculation include the accurate isolation of outflow emission from the disk, dust temperature, dust opacity, and dust-to-gas ratio in the extreme environment of the outflow, and the assumption that the 104\,GHz continuum emission is entirely from dust emission. We would need further data to better constrain these unknowns.

\underline{\it Size:} We measure the size of the lobes by fitting them to an elliptical Gaussian profile. The deconvolved major and minor axis, corrected for inclination, i.e. divided by $\cos(53.5^{\circ})$\footnote{This correction is only applied to the major axis since it is in the north-south direction and then affected by inclination.}, for the north (south) component are 212 mas $\times$ 70 mas (331 mas $\times$ 37 mas), with the solid angle (area) of an elliptical Gaussian being $\mathrm{\pi\times major\times minor/(4ln2)}$.


\underline{\it Age, Mass Loss Rate, and depletion timescale:} Using the inclination-corrected velocities and deconvolved major axes of the lobes we find dynamical ages of $\simeq(0.9-2.3)\times10^5$\,yr, and corresponding mass outflow rates in each lobe of $\dot{M}\geqslant10$\,M$_\odot$\,yr$^{-1}$, depending on whether the values are derived from the CO or HCN emission  (see Table \ref{tbl-1}). 

The dynamical mass of the western nucleus is $\sim$10$^{9}$\,M$_{\odot}$, implying an outflow gas depletion timescale of order $\sim10$\,Myr. The material ejected escapes the western nucleus, but may remain in the observed gas-rich halo \citep[e.g,][]{Sakamoto99}.

\begin{deluxetable*}{lcccc}[th]
\tabletypesize{\scriptsize}
\tablecaption{Properties of the outflow derived from CO (1--0) and HCN (1--0) emission\label{tbl-1}}
\tablewidth{0pt}
\tablehead{\colhead{         } &\colhead{Redshifted/North Component} & \colhead{        }& \colhead{Blueshifted/South Component} & \colhead{ }\\
\colhead{Properties} & \colhead{HCN(1--0)} &\colhead{CO(1--0)} &\colhead{ HCN(1--0)} &\colhead{CO(1--0)}}
\startdata
Velocity (km\,s$^{-1}$) & 856   & 530  & 832   & 825\\
Luminosity  ($\times$10$^{7}$\,K\,km\,s$^{-1}$\,pc$^{2}$)& 2.1   & 0.3   & 0.9  & 1.0\\
M$_{\rm H_{2}}$ (M$_{\odot}$) &  4.9$\times$10$^{6}$ & 2.5$\times$10$^{6}$\tablenotemark{b} & 2.1$\times$10$^{6}$ & 8.0$\times$10$^{6}$\tablenotemark{b} \\
Size lobe (pc$\times$pc) & 78$\times$26 & 112$\times$50  & 122$\times$15 & 190 $\times$30 \\
$\delta$x (pc) & 78 & 112  & 122  & 190\\
Age (10$^{5}$\,yr)  &    0.9    & 2.1  &   1.4  &2.3\\
\.M (M$_{\odot}$ yr$^{-1}$) & 55   &  12   & 15  & 35
\enddata
\tablecomments{}
\tablecomments{All the reported values have been corrected for inclination, i $\sim$ 53.5$^{\circ}$ \citep{BM15}. {\bf Values of Luminosity, M$_{\rm H_{2}}$, and \.M are lower limits}.}
\tablenotetext{b}{Values derived using $\mathrm{\alpha_{CO}= 0.8~M_{\odot} (K~km s^{-1} pc^{2})^{-1}}$.}
\end{deluxetable*}

\section{Discussion}

\subsection{Wind Driving Mechanism}

Taking the total mass outflow rate and velocity to be $\dot{M}=100$\,M$_\odot$\,yr$^{-1}$ and $V=800$\,km\,s$^{-1}$ as reference values (Table \ref{tbl-1}), the total kinetic luminosity and momentum injection rate of the wind are $\dot{E}=\dot{M}V^2/2\sim2\times10^{43}$\,ergs\,s$^{-1}$ and $\dot{P}=\dot{M}V\sim5\times10^{35}$\,dynes, respectively. Both can be accommodated by energy- and momentum-driven wind models discussed in the literature. For example, assuming $10^{51}$\,ergs is injected per SN, per 100\,$M_\odot$ of star formation, SN energy-driven wind models give $V=(2\dot{E}_{\rm hot}/\dot{M}_{\rm hot})^{1/2}\simeq10^3\,{\rm km\,s^{-1}}\,(\alpha/\beta)^{1/2}$, where $\alpha$ is the thermalization efficiency ($\dot{E}_{\rm hot}=\alpha\dot{E}_{\rm SN}$), and $\beta=\dot{M}/{\rm SFR}$ is the hot wind mass loading factor. Given the observed SN rate \citep[e.g.][]{Smith98,Lonsdale06,Varenius17}, and western nucleus ${\rm SFR}\sim100$\,M$_\odot$\,yr$^{-1}$ \citep{BM15} a hot outflow with $\alpha\sim\beta\sim1$ explains its high velocity and accommodates $\dot{M}$. In such a model, the very hot gas $\gtrsim10^7$\,K would be expected to rapidly radiatively cool \citep{Wang95,Thompson16}. Whether it might be able to form molecules in situ, entrain cold gas, or allow the observed dust to survive remains unclear.  

Alternatively, the wind may be driven by radiation pressure on dust. Assuming Keplerian rotation, the observed disk rotation velocity at 60\,pc radius ($\sim280/\sin i$ \,km s$^{-1}$) implies a dynamical mass of $1.1\times10^{9}$\,M$_{\odot}$, and a dust Eddington luminosity of $L_{\rm Edd}=4\pi GMc/\kappa_{\rm IR}\sim10^{12}$\,L$_\odot$, where we assume a Rosseland-mean dust opacity per gram of gas of $\kappa_{\rm IR}=10$\,cm$^2$\,g$^{-1}$. For an ${\rm SFR}\simeq100$\,M$_\odot$\,yr$^{-1}$, the Eddington ratio $\Gamma=L/L_{\rm Edd}\sim1$ \citep{Thompson05}, and one expects a wind \citep{Murray05} with $\dot{P}\sim\tau_{\rm IR}L/c\sim10^{36}$\,dynes\,$(\tau_{\rm IR}/10)(L/10^{12}\,{\rm L_\odot})$ (where $\tau_{\rm IR}$ is the Rosseland-mean optical depth) close to that observed, and with of $V\sim V_{\rm rot}(\Gamma-1)^{1/2}$. Nominally, the latter predicts a maximum velocity below that observed unless $\Gamma\sim6$, but this depends on the distribution of mass and flux (see eq.~5 of \citealt{Zhang&Thompson12}).

Another possibility is that the wind is driven by a central AGN. Assuming efficient energy transfer and conservation would imply an AGN outflow with a total power of the AGN similar to $\dot{E}$ observed \citep[consistent with predicted AGN luminosities based on X-ray observations in][]{Paggi17}. The seemingly collimated nature of the outflow may qualitatively strengthen the argument for AGN driving. Additionally, the finding of a $\sim$20\,pc dust core with a luminosity surface density of $\mathrm{10^{15.5}~L_{\odot}~kpc^{-2}}$ by \citet{Sakamoto17}, may also favor the presence of an AGN.

The true driving mechanism of the outflow is still unclear. We need more complex theoretical models potentially mixing contribution from supernovae and an AGN to predict the outflow presented in this letter.

\subsection{Comparison to other outflows}

Extra-galactic molecular outflows have been detected in CO \citep[e.g.,][]{Alatalo11, Cicone14, Sakamoto14} and HCN \citep[e.g.,][]{Aalto12, GB14, Aalto15, Walter17, Privon17}. Similar outflow velocities and masses have been detected in other systems, however the dynamical time of the outflow presented here is shorter than typically reported values. The smaller spatial scales and the collimation of the outflow reported here are the main differences to those found in starburst galaxies such as NGC\,253 and\,M 82. Indeed, it resembles the molecular jets found towards Galactic protostars, which are signposts for an accreting central object.

\section{Conclusions}
\label{sec:conclusions}

We present the first spatially and spectrally resolved image of the outflow in the western nucleus of Arp\,220. We detect the outflow in HCN~(1--0) thanks to ALMA's  high angular resolution and sensitivity. The wind is also detected in CO(1--0).

The HCN outflow emerges perpendicular to the major axis of the disk and shows maximum inclination-corrected velocities of $\pm850$\,km\,s$^{-1}$ (Table \ref{tbl-1}). The northern and southern lobes exhibit compact and collimated morphology. The mass in these features remains highly uncertain, with an estimated total mass outflow rate of $\geqslant70$\,M$_{\odot}$\,yr$^{-1}$, comparable to the SFR of the western nucleus. The derived characteristics challenge theoretical models of winds driven in extreme starbursts by star formation and/or AGN activity.

\acknowledgments
We thank Viviana Guzm\'an, John Carpenter, Na\'im Ram\'irez-Olivencia, Miguel P\'erez-Torres, Andr\'es P\'erez-Sanchez, and Zhi-Yu Zhang for stimulating discussions. T.A.T. is supported in part by NSF grant \#1516967. K.S. is supported by MOST grant \#106-2119-M-001-025. The work of AKL is partially supported by the National Science Foundation under Grants No. 1615105, 1615109, and 1653300. This research made use of the NASA/IPAC Extragalactic Database (NED), which is operated by the Jet Propulsion Laboratory, California Institute of Technology, under contract with the National Aeronautics and Space Administration, and NASA's Astrophysics Data System Bibliographic Services. The National Radio Astronomy Observatory is a facility of the National Science Foundation operated under cooperative agreement by Associated Universities, Inc. This paper makes use of the following ALMA data: ADS/JAO.ALMA\#2015.1.00702.S and ADS/JAO.ALMA\#2015.1.00113.S. ALMA is a partnership of ESO (representing its member states), NSF (USA) and NINS (Japan), together with NRC (Canada), NSC and ASIAA (Taiwan), and KASI (Republic of Korea), in cooperation with the Republic of Chile. The Joint ALMA Observatory is operated by ESO, AUI/NRAO and NAOJ.

\end{document}